\documentclass[journal]{IEEEtran}
\usepackage[latin1]{inputenc}
\usepackage{amsmath}
\usepackage{graphicx}
\usepackage{amssymb}
\usepackage{xcolor}
\makeatletter

\providecommand{\LyX}{L\kern-.1667em\lower.25em\hbox{Y}\kern-.125emX\@}

\def\BibTeX{{\rm B\kern-.05em{\sc i\kern-.025em b}\kern-.08em
    T\kern-.1667em\lower.7ex\hbox{E}\kern-.125emX}}

\makeatother
\begin{document}
\def\ZZ{{\mathbb Z}}
\def\RR{{\Bbb R}}
\def\NN{{\mathbb N}}
\def\CC{{\mathbb C}}

\title{Delayed Functional Observers for the Realization of Generalized Delayed Control Laws}
	\author{Hieu Trinh
	\thanks{Hieu Trinh is with the School of Engineering, Deakin University, Waurn Ponds, 75 Pigdons Road, Geelong, Australia. (email:  hieu.trinh@deakin.edu.au)}
	}

\maketitle \maketitle \maketitle \thispagestyle{plain}
\pagestyle{plain}

\begin{abstract}

Building on the collective advancements in the literature \cite{trinh1, trinh2, trinhnn26, trinhnam26, trinhnam1}, this paper proposes the design of delayed functional observers to asymptotically estimate a generalized delayed control law under significant input and output delays. This framework enables designers to extend the allowable bounds for input delays while ensuring that the observer-based control scheme stabilizes the system despite simultaneous mismatched input and output time-delays.
\end{abstract}

\begin{keywords}
Time-delay compensators, delayed measurements, input delays, generalized functional observers, observer-based control.
\end{keywords}

\section{System Description and Problem Statement}

We consider the following linear system with mismatched time delays in both control input and output vectors
\begin{align}
	\label{c1}
	\dot{x}(t)&=Ax(t)+Bu(t-\tau_u),\\
		\label{c2}
	y(t)& = Cx(t-\tau_y),
\end{align}
where $x(t)\in \mathbb{R}^n$ is the state vector, $u(t)\in \mathbb{R}^r$ is the control input vector, and $y(t)\in\mathbb{R}^{p}$ is the output vector. The constants $\tau_u>0$ and $\tau_y>0$ represent the time-delays in the input and output channels, respectively. The matrices $A\in\mathbb{R}^{n\times n}$, $B\in\mathbb{R}^{n\times r}$, and $C\in\mathbb{R}^{p\times n}$ are constant. Without loss of generality, it is assumed that $B$ has full column rank and $C$ has full row rank.

This paper investigates the scenario where the output delay dominates the input delay ($\tau_y > \tau_u$) for systems where $A$ is not necessarily Hurwitz. The converse case ($\tau_y \leq \tau_u$) can be readily resolved using the timeline synchronization approach detailed in Scenario 1 of \cite{trinhnn26}. To achieve system stabilization, we employ the generalized delayed control law\begin{align}\label{c3}u(t-\tau_u)&=Fx(t-\tau_u)+F_hx(t-h),\end{align}
where $F, F_h \in \mathbb{R}^{r \times n}$ represent the control gains, and $h > \tau_u$ is an artificial delay introduced as a tuning parameter. Compared to the conventional control law $u(t-\tau_u)=Fx(t-\tau_u)$, the formulation in (\ref{c3}) accommodates a larger allowable delay bound $\tau_u$ while maintaining system stability \cite{trinhnam26}.

However, implementing (\ref{c3}) requires direct knowledge of the delayed functionals $Fx(t-\tau_u)$ and $F_hx(t-h)$. Because neither the current state vector $x(t)$ nor these delayed states are available for feedback, an observer framework is strictly required. Therefore, we propose a design approach to estimate the generalized delayed functional
\begin{equation}
	\label{c4}
	z(t)=Fx(t-\tau_u)+F_hx(t-h).\end{equation}

To realize this control law, this paper develops a systematic framework for various observer configurations. Specifically, we investigate a dual-observer structure that provides a robust and resilient architecture for physical analog hardware deployment. Numerical examples are subsequently presented to validate the theoretical results, ultimately offering the designer a practical range of options for estimating the generalized delayed functional (\ref{c4}).

\section{Dual-observer structure to estimate the generalized functional}
Let us substitute (\ref{c3}) into (\ref{c1}) to obtain the following closed-loop system
\begin{align}
	\label{c5}
	\dot{x}(t)&=Ax(t)+BFu(t-\tau_u)+BF_hx(t-h).
\end{align}

\textit{Case 1 ($h\geq \tau_y$):} We first investigate a dual-observer design to estimate the generalized delayed functional (\ref{c4}) under the condition $h\geq \tau_y$. As demonstrated by Lemma 13 in \cite{trinhnam26}, asymptotic stability of (\ref{c5}) can be maintained even for an extended input delay $\tau_u > \bar{\tau}_u$ under a given $h>\tau_u$, where $\bar{\tau}_u$ denotes the maximum delay allowed by the conventional control law from Lemma 11 \cite{trinhnam26}. This framework introduces a critical design trade-off regarding the parameter $h$. Crucially, within this configuration, $h$ is lower-bounded by the output delay ($h \geq \tau_y$). Thus, the optimal choice of $h$ must balance maximizing the tolerable input delay against satisfying the physical constraints imposed by the output measurement delay.

Now, let
\begin{align}
	\label{c6}
	z(t)=z_1(t)+z_2(t),
\end{align}
where
\begin{align*}
	z_1(t) := Fx(t-\tau_u), \quad z_2(t) := F_hx(t-h), \quad \text{with } h\ge\tau_y.
\end{align*}
Accordingly, the proposed dual-observer scheme employs two separate observers: Observer 1 is designed to estimate the functional $z_1(t)$, while Observer 2 estimates the functional $z_2(t)$. The overall estimate of the generalized delayed functional is then constructed as
\begin{align*}
	\hat{z}(t)=\hat{z}_1(t)+\hat{z}_2(t).
\end{align*}

First, following the framework in \cite{trinh1}, the functional $z_1(t)$ is estimated via a functional observer of the form
\begin{align}
	\label{c7}
	\hat{z}_1(t) &= w_1(t) + M_1y(t), \\
	\label{c8}
	\dot{w}_1(t) &= N_1w_1(t) + N_{1\tau}w_1(t-\tau) + G_1y(t) + G_{1\tau}y(t-\tau) \nonumber \\
	&\quad + Ju(t-2\tau_u) + J_1u(t-\tau_u-\tau_y),
\end{align}
where $\tau = \tau_y - \tau_u > 0$, $w_1(t) \in \mathbb{R}^r$, and the initial condition is given by $w_1(\theta) = \rho(\theta)$ for $\theta \in [-\tau, 0]$. For comprehensive structural details, existence conditions, and the exact parameter determination procedure, the reader is referred to \cite{trinh1}.

Second, since $h \ge \tau_y$, the functional $z_2(t)$ can be estimated using a structurally simpler observer design. Consider the case where a subset of the rows in $F_h$ lies within the row space of $C$. Under this condition, $F_h$ can be decomposed using a transformation matrix $K$ and a matrix $\bar{F}_h$, whose rows are linearly independent of $C$ and are obtained via orthogonal projection. Consequently, the functional $z_2(t)$ can be expressed as
\begin{align}
	\label{c9}
	z_2(t) &=F_hx(t-h)=K\begin{pmatrix} \bar{F}_h\\C \end{pmatrix}x(t-h)\nonumber\\&=K\begin{pmatrix} \bar{z}_2(t)\\y(t-\alpha) \end{pmatrix},
\end{align}
where $\alpha = h - \tau_y\geq 0$, and $\bar{z}_2(t) := \bar{F}_h x(t-h)\in \mathbb{R}^v$ ($0\leq v<r$) represents the independent component requiring observer estimation. Note that given the delayed output measurement $y(t) = Cx(t-\tau_y)$, the relation $y(t-\alpha) = Cx(t-h)$ holds strictly.

Thus, in this scenario, estimating the entire delayed functional is unnecessary. Instead, only $\bar{z}_2(t)$ needs to be estimated, from which the overall estimation of $z_2(t)$ is subsequently obtained as
\begin{align}
	\label{c10}\hat{z}_2(t)=K\begin{pmatrix} \hat{\bar{z}}_2(t)\\y(t-\alpha) \end{pmatrix}.\end{align}

Given that $\tau_y\leq h$, the functional $\bar{z}_2(t)$ can be estimated via the following functional observer structure
\begin{align}
	\label{c11}
	\hat{\bar{z}}_2(t) &= w_2(t) + M_{2\alpha}y(t-\alpha), \\
	\label{c12}
	\dot{w}_2(t) &= N_2w_2(t)+G_{2\alpha}y(t-\alpha) +J_2u(t-h-\tau_u),
\end{align}
where $w_2(t) \in \mathbb{R}^v$. For a comprehensive discussion on the corresponding existence conditions, the reader is referred to Scenario 1 of \cite{trinhnn26}. Additionally, systematic design procedures for calculating these observer parameters can be found in \cite{darouach2000, trinhfer2012}.

\textit{Remark 1 (Observer Order Augmentation):} When designing the functional observer defined by equations (\ref{c7})-(\ref{c8}), the fundamental rank condition$$\text{rank} \begin{pmatrix} FA \\ F \end{pmatrix} = \text{rank}(F)$$may not always hold. In such cases, the order of the observer is increased by augmenting the primary functional $z_1(t) = Fx(t-\tau_u)$ with an auxiliary functional $z_a(t) = Rx(t-\tau_u)$. This yields the augmented functional$$z_{\text{aug}}(t) = \begin{pmatrix} z_1(t) \\ z_a(t) \end{pmatrix} = \begin{pmatrix} F \\ R \end{pmatrix} x(t-\tau_u) = \bar{F}x(t-\tau_u),$$where $\bar{F} := \begin{pmatrix} F \\ R \end{pmatrix} \in \mathbb{R}^{q \times n}$. Following the systematic construction in \cite{trinhnam1}, the matrix $R \in \mathbb{R}^{(q-r) \times n}$ is chosen such that the augmented system satisfies the required rank condition
$$\text{rank} \begin{pmatrix} \bar{F}A \\ \bar{F} \end{pmatrix} = \text{rank}(\bar{F}).$$
Consequently, by designing a $q$-th order functional observer to estimate $z_{\text{aug}}(t)$, the original target estimate $\hat{z}_1(t)$ can be directly obtained.

Furthermore, if $F_h$ lies within the row space of $\bar{F}$, the second functional $z_2(t)$ can be directly realized from $\hat{z}_{\text{aug}}(t)$ as follows
\begin{align}
	\label{c13}
	\hat{z}_2(t)=K_2\hat{z}_{\text{aug}}(t-\gamma),
\end{align}
where $\gamma = h-\tau_u>0$ and $K_2=F_h\bar{F}^-\in\mathbb{R}^{r\times q}$.

In this case, designing a separate observer to estimate $z_2(t)$ is unnecessary; instead, it can be obtained via a linear combination of $\hat{z}_{\text{aug}}(t)$, delayed by $\gamma$.

Extending this rationale, one can also investigate the case where only a subset of the rows in $F_h$ lies within the row space of $\begin{pmatrix} \bar{F} \\ C \end{pmatrix}$. Consequently, the designer only needs to construct a functional observer for the remaining, lower-dimensional sub-functional. This presents a significant architectural advantage, as it reduces the observer's computational complexity while maintaining estimation integrity. The exact implementation details of this reduced-order scheme can be tailored based on the designer's specific system requirements.

The following numerical example serves to clarify the theoretical framework discussed above.

\textit{Example 1:} We consider system (\ref{c1})-(\ref{c2}), where $\tau_u=0.5\text{s}$, $\tau_y=0.95\text{s}$, and matrices $$A=\begin{pmatrix} 1 & 1 \\ -1 & 1 \end{pmatrix}, \quad B=\begin{pmatrix} 0 \\ 1 \end{pmatrix}, \quad C=\begin{pmatrix} 1 & 0 \end{pmatrix}.$$

First, for a delay of $\tau_u = 0.5\text{s}$, the LMI in \cite{trinhnam26} (Lemma 11) is infeasible. Consequently, the conventional control law $u(t-0.5)=Fx(t-0.5)$ fails to stabilize the closed-loop system under this lemma. Second, by employing the generalized delayed control law (\ref{c3}), the LMI in Lemma 13 of \cite{trinhnam26} remains feasible for $\tau_u = 0.5\,\text{s}$ and accommodates an artificial delay parameter of up to $h = 1.3\,\text{s}$, thereby successfully dominating $\tau_y$.

Conversely, if the input delay increases to $\tau_u = 0.52\text{s}$, the LMI in Lemma 13 is only feasible for a maximum artificial delay  of $h = 0.93\text{s}$, which falls short of the physical constraint $\tau_y = 0.95\text{s}$. This variation highlights an inherent design trade-off regarding the parameter $h$, representing a delicate balance between maximizing the system's input-delay tolerance and satisfying the physical constraints imposed by the output measurement delay.

\begin{figure}[!h]
	\centering
	\includegraphics[width=0.9\linewidth]{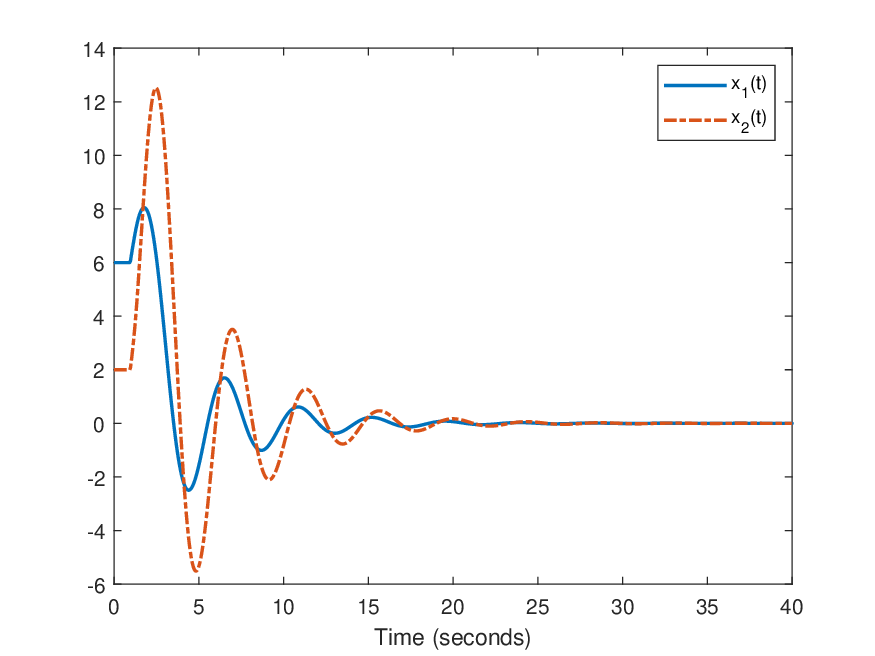}
	\caption{Trajectories of $x_1(t)$ and $x_2(t)$: Generalized delayed control law}
	\label{fig1paper5}
\end{figure}

Now, for $\tau_u=0.5\text{s}$ and by setting $h = \tau_y = 0.95\text{s}$, the observer structure simplifies significantly because $\alpha = h - \tau_y = 0$. This configuration yields the following control law
\begin{align}
	\label{ex}
	u(t-0.5)&=\begin{pmatrix} 0.3288 &  -1.9284 \end{pmatrix}x(t-0.5)\nonumber\\&+\begin{pmatrix} -0.3331  &  0.2068 \end{pmatrix}x(t-0.95).
\end{align}
Figure \ref{fig1paper5} shows the state trajectories of $x_1(t)$ and $x_2(t)$, confirming that asymptotic stability of the closed-loop system is achieved under direct delayed state feedback.

Next, we design an observer to estimate the first functional
\[z_1(t)=\begin{pmatrix} 0.3288  & -1.9284 \end{pmatrix}x(t-0.5)\]
utilizing the observer framework (\ref{c7})-(\ref{c8}). Based on the design procedure \cite{trinhnam26}, we obtain the following second-order functional observer implementation
\begin{align}
	\hat{z}_1(t)& = \begin{pmatrix} 1 & 0\end{pmatrix} \hat{z}_{\text{aug}}(t), \nonumber
\end{align}
where
\begin{align}
\hat{z}_{\text{aug}}(t)&=w_1(t)+\begin{pmatrix} -3.4648\\
	1.8067\end{pmatrix}y(t),\nonumber\\	\dot{\hat{z}}_{\text{aug}}(t)&=\begin{pmatrix}  6.8658 &  11.6406\\-3.0418 &  -4.8658 \end{pmatrix}\hat{z}_{\text{aug}}(t)\nonumber\\&+\begin{pmatrix}-2.4800 &  -1.3177\\
	1.0385 & 0.1960\end{pmatrix}\hat{z}_{\text{aug}}(t-0.45) \nonumber\\&+\begin{pmatrix} 1.5227\\
	-0.4002 \end{pmatrix}y(t)+\begin{pmatrix} 6.2118\\
	-3.2441\end{pmatrix}y(t-0.45)\nonumber\\&+\begin{pmatrix} -1.9284\\
	1 \end{pmatrix}u(t-1), \nonumber
\end{align}
$w_1(t)=\begin{pmatrix} w_{11}(t)\\
	w_{12}(t) \end{pmatrix}$, and corresponding augmented functional given by
\[{z}_{\text{aug}}(t)=\bar{F}x(t-0.5),\quad  \bar{F}=\begin{pmatrix} 0.3288 &   -1.9284\\0 &1\end{pmatrix}.\]

Finally, regarding the second functional, $z_2(t)$, since $F_h$ lies within the row space of $\bar{F}$ (as discussed in Remark 1), the estimate $\hat{z}_2(t)$ can be directly reconstructed from $\hat{z}_{\text{aug}}(t)$ via
$$\hat{z}_2(t)=K_2\hat{z}_{\text{aug}}(t-\gamma),$$
where
\[\gamma = h-\tau_u=0.45\text{s}, \quad K_2=F_h\bar{F}^-=\begin{pmatrix} -1.0132 &  -1.7471
\end{pmatrix}.\]

\begin{figure}[!h]
	\centering
	\includegraphics[width=0.9\linewidth]{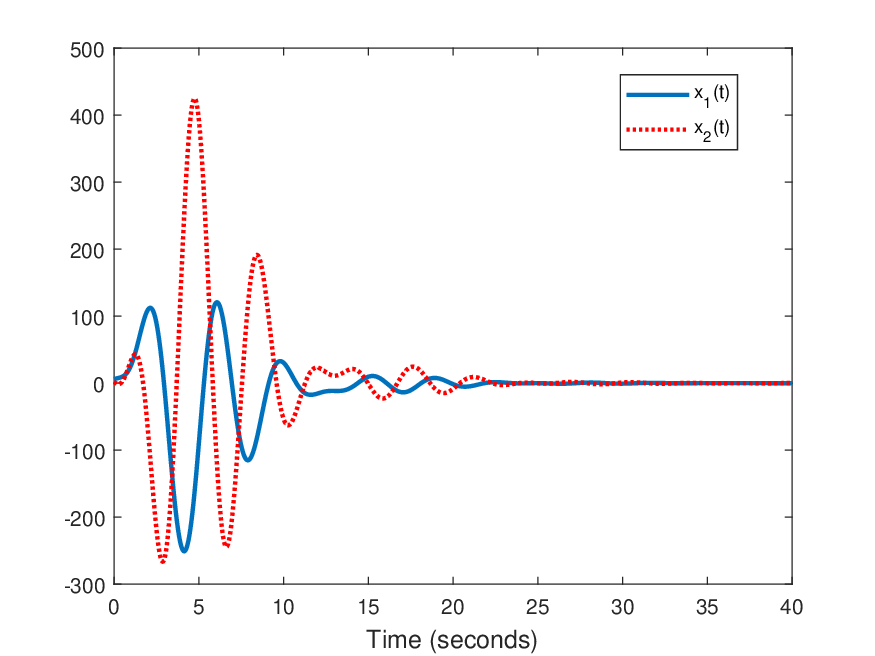}
	\caption{Trajectories of $x_1(t)$ and $x_2(t)$: Observer-based control with $R=\begin{pmatrix} 0 & 1 \end{pmatrix}$}
	\label{fig2paper5}
\end{figure}

By integrating the designed controller and observer, the output feedback closed-loop system achieves asymptotic stability. This is verified by the trajectories of $x_1(t)$ and $x_2(t)$ in Figure \ref{fig2paper5}, where all states successfully converge to origin. Notably, although the transient responses exhibit high initial peaks---primarily due to the observer's peaking phenomenon---they subsequently decay rapidly to zero.

\textit{Remark 2 (Suppression of high peaks):} As observed in Figure \ref{fig2paper5}, the transient response of the observer-based control scheme exhibits severe peaks. Minimizing these transient peaks is of significant practical importance. Crucially, the proposed augmentation approach (see Remark 1) allows the designer to exploit the inherent freedom in selecting the additional matrix $R$. For instance, instead of choosing $R = \begin{pmatrix} 0 & 1 \end{pmatrix}$ as in Example 1, we select
$$R = \begin{pmatrix} 5 & 0 \end{pmatrix}$$
which yields
$$\bar{F} = \begin{pmatrix} 0.3288 & -1.9284 \\ 5 & 0 \end{pmatrix}, \quad K_2 = \begin{pmatrix} -0.1072  & -0.0596 \end{pmatrix}.$$

Then, following the design procedure in \cite{trinhnam26}, we construct a second-order functional observer to estimate the augmented functional
$$z_{\text{aug}}(t) = \bar{F}x(t-0.5)$$from which the desired estimates are extracted via
$$\hat{z}_1(t) = \begin{pmatrix} 1 & 0 \end{pmatrix} \hat{z}_{\text{aug}}(t), \ \hat{z}_2(t)=K_2\hat{z}_{\text{aug}}(t-\gamma).$$

\begin{figure}[!h]
	\centering
	\includegraphics[width=0.9\linewidth]{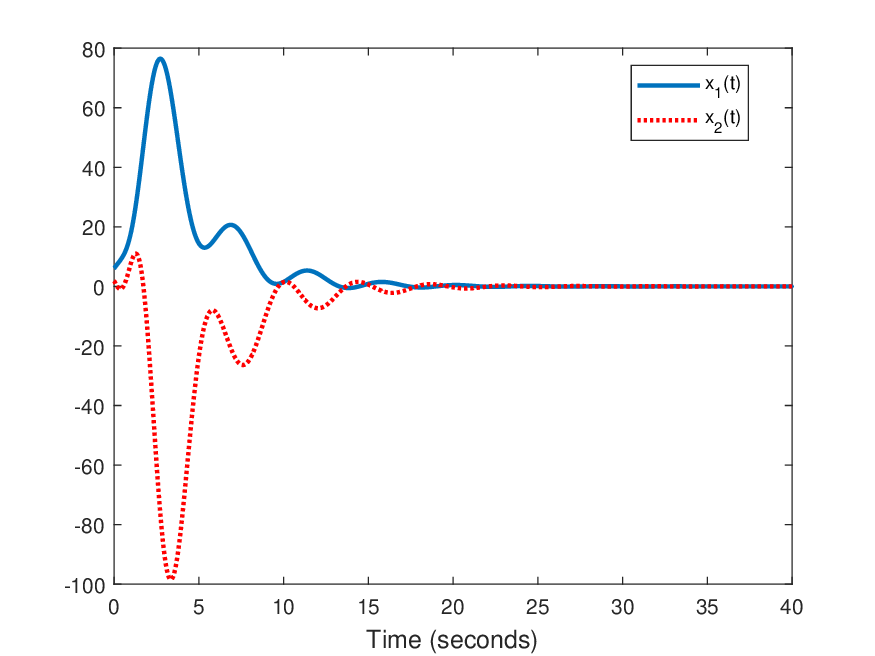}
	\caption{Trajectories of $x_1(t)$ and $x_2(t)$: Observer-based control with $R=\begin{pmatrix} 5 & 0 \end{pmatrix}$}
	\label{fig3paper5}
\end{figure}

The design details are omitted here for brevity. Figure \ref{fig3paper5} illustrates the resulting closed-loop responses of the observer-based control system. Notably, the large transient peaks are significantly reduced compared to those in Figure \ref{fig2paper5}. This demonstrates that the design freedom in selecting $R$ can be effectively leveraged to satisfy the rank condition (i.e., $\text{rank} \begin{pmatrix} \bar{F}A \\ \bar{F} \end{pmatrix} = \text{rank}(\bar{F})$) while simultaneously mitigating transient issues. However, a systematic methodology for optimally selecting $R$ to minimize peak overshoots remains an open problem. At present, a trial-and-error or fine-tuning approach is required to achieve the desired reduction, leaving the development of a systematic selection framework as an open direction for future research.

\textit{Case 2 ($h < \tau_y$):} In the remainder of this section, we investigate a dual-observer design to estimate the generalized delayed functional (\ref{c4}) under the condition $h < \tau_y$. This situation arises either deliberately to expand $\tau_u$ beyond the upper bound $\bar{\tau}_u$ (resulting in a reduction of $h$, as discussed in Example 1), or inherently due to physical constraints imposed on the output measurement delay. 

In either case, the estimation is first carried out by using a primary observer to estimate the intermediate state functional $z_1(t) = Fx(t-\tau_u)$. This initial step can be realized via a functional observer of the form (\ref{c7})-(\ref{c8}) with the net delay parameter $\tau = \tau_y - \tau_u$. However, if this observer cannot ensure asymptotic stability of the error system, the multi-delay functional observer configuration from \cite{trinh1} (specifically, equations (11)-(12) therein) can be employed instead. By incorporating multiple internal delays to expand $\tau$, this alternative formulation improves the robustness of the error-estimation time-delay system. Ultimately, the architecture in \cite{trinh1} offers a deliberate design trade-off, trading increased parameter complexity for enhanced estimation robustness.

Next, because $\tau_y > h$, the observer structure (\ref{c11})--(\ref{c12}) cannot be employed to estimate the second functional $z_2(t) = F_hx(t-h)$ since the parameter $\alpha = h - \tau_y < 0$. To circumvent this limitation, we leverage the framework established in Remark 1. Assuming the target estimate $\hat{z}_1(t)$ is extracted from the augmented estimate $\hat{z}_{\text{aug}}(t)$, we examine the scenario where either $F_h$ or a subset of its rows lies within the row space of $\bar{F}$. Under these conditions, the following structural simplifications are achieved:

\textit{(i) $F_h$ lies within the row space of $\bar{F}$:} Under this condition, the second functional $z_2(t)$ can be directly reconstructed from $\hat{z}_{\text{aug}}(t)$ according to equation (\ref{c13}). Consequently, designing a separate observer to estimate $z_2(t)$ becomes redundant; it is obtained instead via a linear combination of $\hat{z}_{\text{aug}}(t)$ delayed by $\gamma$.

\textit{(ii) A subset of the rows of $F_h$ lies within the row space of $\bar{F}$:} Under this condition, $F_h$ can be decomposed using a transformation matrix $K$ and a matrix $\bar{F}_h$, whose rows are linearly independent of $\bar{F}$ and are obtained via orthogonal projection. Consequently, the functional $z_2(t)$ can be expressed as
\begin{align}
	\label{c15}
	z_2(t) &=F_hx(t-h)=K\begin{pmatrix} \bar{F}_h\\\bar{F} \end{pmatrix}x(t-h)\nonumber\\&=K\begin{pmatrix} \bar{z}_2(t)\\\hat{z}_{\text{aug}}(t-\gamma) \end{pmatrix},
\end{align}
where $\gamma = h - \tau_u>0$, and $\bar{z}_2(t) := \bar{F}_h x(t-h)\in \mathbb{R}^v$ ($0<v<r$) represents the independent component requiring observer estimation. 
Accordingly, the overall estimate is constructed as
\begin{align}
	\label{c17}
	\hat{z}_2(t)=K\begin{pmatrix} \hat{\bar{z}}_2(t)\\ \hat{z}_{\text{aug}}(t-\gamma) \end{pmatrix},
\end{align}
where the estimation of the lower-dimensional sub-functional $\bar{z}_2(t)$ is carried out using the observer structure (\ref{c7})--(\ref{c8}). Furthermore, the asymptotic stability requirement of the resulting error time-delay system is structurally less conservative. This is because $h > \tau_u$ implies that the net time delay $\tau = \tau_y - h$ is strictly smaller than the delay encountered in the primary functional observer, where $\tau = \tau_y - \tau_u$.

\begin{figure}[!h]
	\centering
	\includegraphics[width=0.9\linewidth]{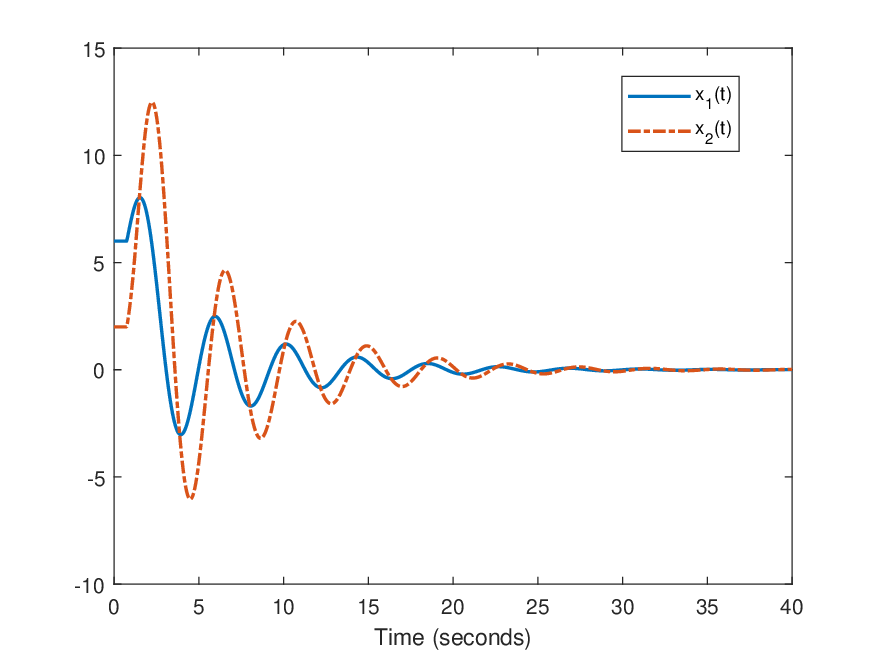}
	\caption{Trajectories of $x_1(t)$ and $x_2(t)$: Generalized delayed control law}
	\label{fig4paper5}
\end{figure}

\textit{Example 2:} To illustrate Case 2, let us reconsider Example 1, where the system delays are given as $\tau_u = 0.53\text{s} > \bar{\tau}_u$ and $\tau_y = 1\text{s}$. Given this value of $\tau_u$, the LMI condition provided in \cite{trinhnam26} (Lemma 13) is feasible only up to a maximum delay margin of $h = 0.81\text{s}$. Crucially, this upper bound falls short of the system's output delay $\tau_y = 1\text{s}$. By choosing the design parameter $h = 0.75\text{s} < 0.81\text{s}$ under the given $\tau_u = 0.53\text{s}$, the LMI in Lemma 13 in  \cite{trinhnam26} is feasible with $\lambda = 1.05$, yielding the following control law\begin{align}
	u(t-0.53)&=\begin{pmatrix} 0.2518  & -2.3939 \end{pmatrix}x(t-0.53)\nonumber\\&+\begin{pmatrix} -0.2247 &   0.7191 \end{pmatrix}x(t-0.75).\nonumber
\end{align}
Figure \ref{fig4paper5} shows the trajectories of $x_1(t)$ and  $x_2(t)$. It is clear
that asymptotic stability of the closed-loop system has been achieved.

Next, we design an observer to estimate the first functional
\[z_1(t)=u(t-0.53)=\begin{pmatrix} 0.2518  & -2.3939 \end{pmatrix}x(t-0.53)\]
using the observer (\ref{c7})-(\ref{c8}). Based on the design procedure \cite{trinhnam26}, we obtain the following second-order functional observer
\begin{align}
	\hat{z}_1(t)& = \begin{pmatrix} 1 & 0\end{pmatrix} \hat{z}_{\text{aug}}(t), \nonumber
\end{align}
where
\begin{align}
	\hat{z}_{\text{aug}}(t)&=w_1(t)+\begin{pmatrix} -2.5003\\
		2.7373\end{pmatrix}y(t),\nonumber\\	\dot{\hat{z}}_{\text{aug}}(t)&=\begin{pmatrix}  0.8948 &    0.4841\\-2.0886 &   1.1052 \end{pmatrix}\hat{z}_{\text{aug}}(t)\nonumber\\&+\begin{pmatrix}-1.0445 &  -0.3680\\1.1435 &  -1.6119\end{pmatrix}\hat{z}_{\text{aug}}(t-0.47) \nonumber\\&+\begin{pmatrix} 3.6910\\
		13.2821 \end{pmatrix}y(t)+\begin{pmatrix} 1.6042\\
		-7.2715\end{pmatrix}y(t-0.47)\nonumber\\&+\begin{pmatrix} -2.3939\\
		0 \end{pmatrix}u(t-1.06), \nonumber
\end{align}
$w_1(t)=\begin{pmatrix} w_{11}(t)\\
	w_{12}(t) \end{pmatrix}$, and the augmented functional
\[{z}_{\text{aug}}(t)=\bar{F}x(t-0.53),\quad  \bar{F}=\begin{pmatrix} 0.2518  & -2.3939\\5 &0\end{pmatrix}.\]

Finally, regarding the second functional, $z_2(t)$, since $F_h$ lies within the row space of $\bar{F}$, $\hat{z}_2(t)$ can be reconstructed from $\hat{z}_{\text{aug}}(t)$ as follows
$$\hat{z}_2(t)=K_2\hat{z}_{\text{aug}}(t-\gamma),$$
where
\[\gamma = h-\tau_u=0.22\text{s}, \quad K_2=F_h\bar{F}^-=\begin{pmatrix} -0.3004  & -0.0298\end{pmatrix}.\]

\begin{figure}[!h]
	\centering
	\includegraphics[width=0.9\linewidth]{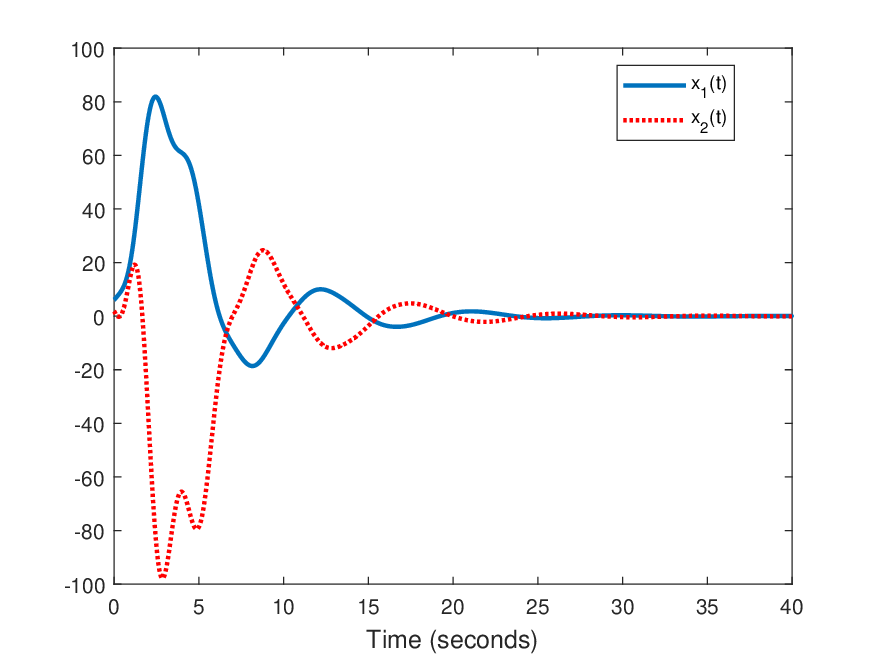}
	\caption{Trajectories of $x_1(t)$ and $x_2(t)$: Observer-based control with $R=\begin{pmatrix} 5 & 0 \end{pmatrix}$, $\tau_y>h$}
	\label{fig5paper5}
\end{figure}

By integrating the designed controller and observer, the closed-loop system achieves asymptotic stability. This is confirmed by the trajectories of $x_1(t)$ and $x_2(t)$ in Figure \ref{fig5paper5}, where all states successfully converge to zero.

\section{Conclusion} This paper expands upon the existing literature \cite{trinh1, trinh2, trinhnn26, trinhnam26, trinhnam1} by developing a robust functional observer framework tailored for systems with significant, mismatched input and output delays. The proposed method guarantees the asymptotic estimation of a generalized delayed control law, ultimately providing control engineers with a reliable strategy to maximize allowable input delay thresholds without sacrificing overall system stability.


\begin{thebibliography}{99}
	\bibitem{trinh1} H. Trinh, ``Delayed functional observers for output-delayed linear systems'', Preprint at 
	https://doi.org/10.48550/arXiv.2606.09407 (2026).
	
	\bibitem{trinh2} H. Trinh, ``On time-delay compensators for delayed-output systems'', Preprint at 
	
	https://doi.org/10.48550/arXiv.2606.10308 (2026).
	
	\bibitem{trinhnn26} H. Trinh, P. T. Nam and T. N. Nguyen, ``Observer-based control of linear systems with mismatched input and output delays'', Preprint at 
	
	https://doi.org/10.48550/arXiv.2606.03081 (2026).
	\bibitem{trinhnam26} H. Trinh, P. T. Nam and T. N. Nguyen, ``Time-Delay compensators for linear systems with delayed output measurements'', Preprint at 
	https://doi.org/10.48550/arXiv.2604.17434 (2026).
	
	\bibitem{trinhnam1} H. Trinh, P. T. Nam and T. Fernando, ``Existence and design of functional observers for time-delay systems with delayed output measurements'', Preprint at 
	https://doi.org/10.48550/arXiv.2603.09395 (2026).
	
	\bibitem{darouach2000}  M. Darouach, ``Existence and design of functional observers'', {\it IEEE Trans. Autom. Contr.}, vol. 45, no.5, pp. 940-943, 2000.
	
	\bibitem{trinhfer2012} H. Trinh and T. Fernando, \textit{Functional Observers for Dynamical Systems}. Springer-Verlag,
	Berlin Heidelberg, 2012.


\end{thebibliography}
\end{document}